\documentclass[10pt,twocolumn]{article} 

\let\cmtt=\ttdefault
\usepackage{latex8}
\usepackage{times}
\let\ttdefault=\cmtt
\usepackage[override]{cmtt}

\usepackage[final]{showkeys}

\usepackage{fancyvrb}
\usepackage{wrapfig}
\usepackage{amsmath}
\usepackage{amssymb}
\usepackage{amsfonts}
\usepackage{color}
\usepackage{url}
\usepackage{tikz}
\usepackage[vlined]{algorithm2e}
\SetKwRepeat{Repeat}{do}{while}

\newtheorem{theorem}{Theorem}[section]

\newtheorem{defi}[theorem]{Definition}
\newtheorem{rema}[theorem]{Remark}

\newtheorem{exam}[theorem]{Example}

\usepackage{paralist}

\usepackage{nicefrac}

\usepackage{xml}

\makeatletter
\newcommand{\namelabel}[2]{\def\@currentlabel{#1}\label{#2}}
\makeatother

\newcommand{\ds}[2]{\fun{ds}_{#1}\left(#2\right)}
\newcommand{\distlabel}[2]{\namelabel{\textsc{#1}}{#2}(\ref{#2})}

\usepackage{mathpartir}

\usetikzlibrary{arrows}
\usetikzlibrary{snakes}

\newcommand{\bigstep}[4]{%
  \begin{tikzpicture}
    \node at (-2ex,0ex) [minimum width=#3,
                         minimum height=2ex,
                         white,fill=black!20,
                         inner sep=.5pt,outer sep=0pt,
                         rotate=90,anchor=north west] (n) 
                         {\scriptsize\color{black!80}#1};
    \draw[densely dashed,black!50] (n.north east) rectangle (#2,0ex) + (n.south west);
    \path (n.south east) -- (#2,0ex) + (n.south west) node[midway] {#4};
  \end{tikzpicture}
}

\newcommand{\aplan}[1]{%
  \begin{tikzpicture}[x=2.5mm,y=3.5mm,line width=0.3pt]
    \draw[rounded corners=1pt]
      ++(-1,-0.5) --  ++(2,0) -- ++(-1,1.5) -- cycle;
    \node {\scriptsize#1};
  \end{tikzpicture}
}

\newcommand{\waplan}[1]{%
  \begin{tikzpicture}[x=4.5mm,y=4mm,line width=0.3pt]
    \draw[rounded corners=1pt]
      ++(-1,-0.5) --  ++(2,0) -- ++(-1,1.5) -- cycle;
    \node {\scriptsize#1};
  \end{tikzpicture}
}

\newcommand{\miniaplan}[1]{%
  \begin{tikzpicture}[baseline=-0.5mm,x=2mm,y=2.8mm,line width=0.2pt]
    \draw[rounded corners=1pt]
      ++(-1,-0.5) --  ++(2,0) -- ++(-1,1.5) -- cycle;
    \node {\scriptsize#1};
  \end{tikzpicture}
}

\DeclareMathOperator{\recin}{%
  \begin{tikzpicture}[baseline=0.1ex,scale=0.55,x=1mm,y=4.5mm,inner sep=1pt]
    \draw[)-] (0,0) -- (0,1);
    \useasboundingbox (-2,0) rectangle (2,1);
  \end{tikzpicture}
}

\DeclareMathOperator{\recout}{%
  \begin{tikzpicture}[baseline=0.1ex,scale=0.55,x=1mm,y=4.5mm,inner sep=1pt]
    \draw[-o] (0,0) -- (0,1);
    \useasboundingbox (-2,0) rectangle (2,1);
  \end{tikzpicture}
}

\DeclareMathOperator{\scj}{%
  \begin{tikzpicture}[scale=0.085,baseline=0.1ex,join=round,x=1cm,y=1cm]
    \draw (0,0) -- (0,1) -- (1,1) -- (1,2) -- (2,2) -- (2,3) -- (3,3) --
          (3,0) -- cycle;
    \useasboundingbox (0,0) rectangle (3.5,3);        
  \end{tikzpicture}
}

\newcommand{\num}{\text{\footnotesize\#}}

\newcommand{\col}[1]{\mathsf{#1}}

\usepackage{colortbl}
\newcolumntype{H}{>{\columncolor{black}\color{white}}c}
\newcommand{\colhd}[1]{\multicolumn{1}{H}{\col{#1}}}
\newenvironment{littbl}{%
  \renewcommand{\arraystretch}{0.7}
  }{}

\usepackage{booktabs}

\usepackage[tight,small,raggedright]{subfigure}

\usepackage{stmaryrd}

\usepackage{xspace}

\newcommand{\naive}{na\"ive\xspace}
\newcommand{\Naive}{\emph{Na\"ive}\xspace}
\newcommand{\DELTA}{\emph{Delta}\xspace}

\newcommand{\fv}[1]{\fun{fv}(#1)}

\usepackage{overlay}
\newcommand{\setequal}{%
  \mathbin{\smash[t]{\overset{s}{=}}}}
\newcommand{\nsetequal}{%
  \mathbin{\ontopof{\smash[t]{\overset{s}{\phantom{=}}}}{\neq}}}

\begin{document}
\VerbatimFootnotes

\title{%
  \vspace*{-5mm}%
  An Inflationary Fixed Point Operator in XQuery}

\newcommand{\TUM}{$\,^{\displaystyle\circ}$}
\newcommand{\UVA}{$\,^{\displaystyle\bullet}$}

\author{%
  \begin{tabular*}{\textwidth}{@{\extracolsep\fill}ccccc}
  Loredana Afanasiev\UVA
  &
  Torsten Grust\TUM
  &
  Maarten Marx\UVA
  &
  Jan Rittinger\TUM  &
  Jens Teubner\TUM
\end{tabular*}
}

\affiliation{%
  \UVA{}\,ISLA, University of Amsterdam, Amsterdam, The Netherlands \\
  \TUM{}\,Technische Universit\"at M\"unchen, Munich, Germany 
}

\email{%
  \url|{lafanasi,marx}@science.uva.nl|\textnormal, 
  \url|{grust,rittinge,teubnerj}@in.tum.de| 
}

\maketitle

\begin{abstract}
  We introduce a controlled form of recursion in XQuery,
  \emph{inflationary fixed points}, familiar in the context of
  relational databases. This imposes restrictions on the
  expressible types of recursion, but we show that inflationary fixed points
  nevertheless are sufficiently versatile to capture a wide range of
  interesting use cases, including the semantics of Regular~XPath and
  its core transitive closure construct.
  
  While the optimization of general user-defined recursive functions
  in XQuery appears elusive, we will describe how inflationary fixed points
  can be efficiently evaluated, provided that the recursive XQuery 
  expressions exhibit a \emph{distributivity} property.  We show
  how distributivity can be assessed both, syntactically and algebraically,
  and provide experimental evidence that XQuery processors can
  substantially benefit during inflationary fixed point evaluation.
\end{abstract}

\Section{Introduction}

\newenvironment{definition}[1]{%
  \begin{defi}\textbf{\itshape #1}\normalfont}{\hfill $\lhd$\end{defi}}
\newenvironment{example}{\begin{exam}\normalfont}{\hfill$\lhd$\end{exam}}
\newenvironment{remark}{\begin{rema}\normalfont}{\end{rema}}
\newenvironment{myproof}{\begin{trivlist}\item[]\rm{\bf
Proof.}\em}{\hfill {\sc qed}\end{trivlist}}
\newenvironment{myproofof}[1]{\begin{trivlist}\item[\hskip\labelsep{\bf
Proof~of~{#1}.\ }]}{\hspace*{\fill} {\sc qed}\end{trivlist}}

The backbone of the XML data model, namely \emph{ordered, unranked trees
of nodes}, is inherently recursive and it is natural to equip the
associated languages with constructs that can query such recursive
structures. To get from the recursive axes in XPath, \emph{e.g.},
\xquery{ancestor} and \xquery{descendant}, to XQuery's~\cite{XQuery}
recursive user-defined functions, language designers took a giant leap,
however. User-defined functions in XQuery admit \emph{arbitrary} types
of recursion---a construct that largely evades optimization approaches
beyond ``procedural'' improvements like tail-recursion elimination or
unfolding.

This paper embarks on a journey that explores a controlled form of
recursion in XQuery, the \emph{inflationary fixed point (IFP)}, familiar
in the context of relational databases \cite{dbbook}. While this imposes
restrictions on the expressible types of recursion, IFP embraces a
family of widespread use cases of recursion in XQuery, including many
forms of horizontal or vertical structural recursion and the pervasive
\emph{transitive closure} problem (IFP captures \emph{Regular XPath}
\cite{BtCRegPath:06}, in particular).
\begin{example}\label{ex:curriculum}
  The DTD of Figure~\ref{fig:curri-data} (taken from \cite{xlinkit})
  describes recursive curriculum data, including courses, their lists of
  prerequisite courses, the prerequisites of the latter, and so on. The
  XQuery program of Figure~\ref{fig:fix-intro} uses the \xquery{course}
  element node with code \xquery{"c1"} to seed a computation that
  recursively finds all prerequisite courses, direct or indirect, of
  course \xquery{"c1"}. For a given sequence \xquery{\$x} of
  \xquery{course} nodes, function \xquery{fix($\cdot$)} calls out to
  \xquery{rec($\cdot$)} to find their prerequisites. While new nodes are
  encountered, \xquery{fix($\cdot$)} calls itself with the accumulated
  \xquery{course} node sequence. (This is not expressible in XPath~2.0.)
\end{example}
Note that \xquery{fix($\cdot$)} implements a generic fixed point
computation: only the initialization (\xquery{let \$seed := $\cdots$})
and the \emph{payload function} \xquery{rec($\cdot$)} are specific to the
curriculum problem.  This motivates the introduction of a syntactic
form that can succinctly accommodate this pattern of computation 
(Section~\ref{sec:IFP}).

\begin{figure}
  \centering\small
\begin{BVerbatim}[baselinestretch=0.9]
<!ELEMENT curriculum (course)*>
<!ELEMENT course prerequisites>
<!ATTLIST course code ID #REQUIRED>
<!ELEMENT prerequisites (pre_code)*>
<!ELEMENT pre_code #PCDATA>
\end{BVerbatim}
\caption{Curriculum data (simplified DTD).}
\label{fig:curri-data}
\end{figure}

\begin{figure}
  \centering\small
  \newsavebox{\dashedbox}
  \savebox{\dashedbox}{%
    \tikz[baseline=0.2ex]\draw[densely dashed] (0,0) rectangle (0.7,0.3);}
  \begin{tikzpicture}[overlay]
    \draw[densely dashed] (-3.3,-1.6) rectangle (2.9,-3.75);
  \end{tikzpicture}
  \renewcommand{\theFancyVerbLine}{%
    \scriptsize\normalfont\arabic{FancyVerbLine}}
  \begin{Verbatim}[numbers=left,xleftmargin=1cm,baselinestretch=0.9,
                   commandchars=\\\{\},codes={\catcode`$=3\catcode`\_=8}]
declare function rec\,(\$cs) as node()* 
\{ \$cs/id\,(./prerequisites/pre\_code)
\};

declare function fix\,(\$x) as node()* 
\{ let \$res := rec\,(\$x)
  return if (empty\,(\$x except \$res))
         then \$res
         else fix\,(\$res union \$x)
\};

let \$seed := doc\,("curriculum.xml")
                      /course[@code="c1"]
return fix\,(rec\,(\$seed))
  \end{Verbatim}
  \caption{Prerequisites for the course \xquery{"c1"} 
    (\,\usebox{\dashedbox} marks the fixed point computation).}
  \label{fig:fix-intro}
\end{figure}

Most importantly, however, such computation in IFP form is susceptible
to systematic optimization, provided that the payload (or \emph{body})
of the recursion exhibits a specific \emph{distributivity} property.

Unlike the general user-defined XQuery functions, this account of
recursion puts the query processor into control in that it can decide
whether the optimization may be safely applied. Distributivity may be
assessed on a syntactical level---a non-invasive approach that can
easily be realized on top of existing XQuery processors
(Section~\ref{sec:trading}). Further, though, if we adopt a relational view of
the XQuery semantics (as in \cite{sql-mapping}), the seemingly
XQuery-specific distributivity notion turns out to be elegantly and
uniformly tractable on the familiar algebraic level
(Section~\ref{sec:dist_impl}).

Compliance with the restriction that IFP imposes on query formulation is
rewarded by significant query runtime savings that the IFP-inherent
optimization hook can offer. We document the effect for the XQuery
processors \emph{MonetDB/XQuery} \cite{monetdb-xquery} and \emph{Saxon}
\cite{saxon} in Section~\ref{sec:evaluation}. This is primarily due to a
substantial reduction of the number of items that are fed into the
recursion's payload function (the \naive implementation of
Example~\ref{ex:curriculum} feeds already discovered \xquery{course}
element nodes back into \xquery{rec($\cdot$)}).

In Section~\ref{sec:related}, we stop by related work on recursion on
the XQuery as well as the relational side of the fence, and finally
wrap-up in Section~\ref{sec:wrap-up}.

\Section{An Inflationary Fixed Point in XQuery}
\label{sec:IFP}

\renewenvironment{definition}[1]{%
  \begin{defi}\textbf{\itshape #1}\normalfont}{\hfill $\lhd$\end{defi}}
\renewenvironment{example}{\begin{exam}\normalfont}{\hfill$\lhd$\end{exam}}
\renewenvironment{remark}{\begin{rema}\normalfont}{\end{rema}}
\renewenvironment{myproof}{\begin{trivlist}\item[]\rm{\bf
Proof.}\em}{\hfill {\sc qed}\end{trivlist}}
\renewenvironment{myproofof}[1]{\begin{trivlist}\item[\hskip\labelsep{\bf
Proof~of~{#1}.\ }]}{\hspace*{\fill} {\sc qed}\end{trivlist}}

The subsequent discussion will revolve around the recursion pattern
embodied by function \xquery{fix($\cdot$)} of
Figure~\ref{fig:fix-intro}, known as the \emph{inflationary fixed point
(IFP)} \cite{dbbook}. We will introduce a new syntactic form to
introduce IFP on the XQuery language level and then explore its
semantics in the XQuery context, application, and optimization.

In the following, we regard an XQuery expression $e_1$
containing a free variable \xquery{\$x} as a function of \xquery{\$x}.
We write $e_1(e_2)$ to denote $e_1[\nicefrac{\displaystyle
e_2}{\displaystyle\xquery{\$x}}]$, \emph{i.e.}, the consistent
replacement of all free occurrences of \xquery{\$x} in $e_1$ by $e_2$.
Function $\fv{e}$ returns the set of free variables of expression $e$.
We further introduce \emph{set-equality} ($\setequal$), a relaxed notion
of equality for XQuery item sequences that disregards
duplicate items and order, \emph{e.g.}, $\xquery{(1,"a")} \setequal
\xquery{("a",1,1)}$. 

To streamline the discussion, in the following we assume computations
over sequences of type \xquery{node()*} as trees are \emph{the} recursive data structure
in the XQuery Data~Model.  In this case, with $X_1$, $X_2$ of type
\xquery{node()*}, we have\footnote{Here and in the following,
\xquery{fs:ddo($\cdot$)} abbreviates the function
\xquery{fs:distinct-doc-order($\cdot$)} of the XQuery Formal Semantics
\cite{xquery_fs}.}
$$ 
X_1 \setequal X_2
\quad\Leftrightarrow\quad 
\xquery{fs:ddo($X_1$)} = \xquery{fs:ddo($X_2$)} \enskip. 
$$ 
An extension to general sequences of type
\xquery{item()*} is possible and entails the replacement of XQuery's
node set operations (\xquery{union}, \xquery{except}) with appropriate
variants.

\begin{definition}{(Inflationary Fixed Point)}
\label{def:ifp}
Let $e_{\var{seed}}$ and $e_{\var{rec}}(\xquery{\$x})$ be XQuery
expressions of type \xquery{node()*}. The \emph{inflationary fixed point (IFP) of
$e_{\var{rec}}(\textnormal{\xquery{\$x}})$ seeded by $e_{\var{seed}}$} is an XQuery
expression represented by the following syntactic form:
\begin{equation}
\xquery{with \$x seeded by }e_{\var{seed}}\xquery{ recurse }e_{\var{rec}}(\xquery{\$x})
\enskip .
\label{xq:ifp}
\end{equation}
The payload expression $e_{\var{rec}}$ is called the \emph{body},
$e_{\var{seed}}$ is called the \emph{seed}, and $\xquery{\$x}$ is
called the \emph{recursion variable} of the inflationary fixed point
operator.

The semantics of the IFP of $e_{\var{rec}}(\xquery{\$x})$ seeded by
$e_\var{seed}$ is the sequence of nodes $\var{res}_k$, if it exists,
obtained in the following manner:
$$
\begin{array}{lcl}
  \var{res}_0     & \gets & e_{\var{rec}}(e_{\var{seed}}) \\
  \var{res}_{i+1} & \gets & 
    e_{\var{rec}}(\var{res}_i) \mathbin{\xquery{union}} \var{res}_i
    \quad,\quad i \geqslant 0
\end{array}
$$
where $k \geqslant 1$ is the minimum number for which $\var{res}_{k}
\setequal \var{res}_{k-1}$. Otherwise, the IFP of
$e_{\var{rec}}(\xquery{\$x})$ seeded by $e_{\var{seed}}$ is \emph{undefined}. 
\end{definition}
Note that if expression $e_{\var{rec}}$ does \emph{not} invoke node constructors 
(\emph{e.g.}, \xquery{element\,\{$\cdot$\}\,\{$\cdot$\}} or \xquery{text\,\{$\cdot$\}}), 
such that the query operates over a finite domain of nodes, IFP will 
always be defined.  Otherwise, the invocation of node constructors in the 
recursion body might yield an infinite node domain and IFP might be undefined.

\begin{example}
In terms of the new $\xquery{with}\cdots\xquery{seeded by}\cdots$
$\xquery{recurse}$ 
syntactic form, we can now express the transitive closure query from
Example~\ref{ex:curriculum} in a quite concise and elegant fashion:
\begin{equation}
\begin{BVerbatim}[baseline=c,commandchars=\\\{\},codes={\catcode`$=3\catcode`\_=8}]
with \$x seeded by doc\,("curriculum.xml")
                    /course[@code="c1"]
recurse \$x/id\,(./prerequisites/pre\_code)
\end{BVerbatim}
\tag{Q1}
\label{q:curri}
\end{equation}
\end{example}


\smallskip\noindent
Obviously, the new form $\xquery{with}\cdots\xquery{seeded
by}\cdots\xquery{recurse}$ is mere syntactic sugar as it can be
equivalently expressed via the recursive user-defined function template
\xquery{fix($\cdot$)} (shown in \tikz[baseline=0.2ex]\draw[densely
dashed] (0,0) rectangle (0.7,0.3); in Figure~\ref{fig:fix-intro}). Since
the syntactic form is a second-order construct taking an XQuery variable
name and two XQuery expressions as arguments, function
\xquery{fix($\cdot$)} has to be interpreted as a template in which the
recursion body \xquery{rec($\cdot$)} needs to be instantiated (XQuery~1.0 does not
support higher-order functions). Given this, Expression~\eqref{xq:ifp} is
equivalent to the expression
\begin{BVerbatim}[commandchars=\\\{\},codes={\catcode`$=3\catcode`\_=8}]
let \$x := $e_{\var{seed}}$ return fix\,(rec\,(\$x))
\end{BVerbatim} 
.

 
\paragraph{Using IFP to Compute Transitive Closure.} Much like in the
relational context, \emph{transitive closure} is an archetype of
recursive computation over XML instances. Regular~XPath
\cite{BtCRegPath:06}, for example, defines the transitive closure of XPath
location steps to obtain powerful primitives that express horizontal and
vertical structural recursion. We can naturally extend this definition
to any XQuery expression of type \xquery{node()*}.
\begin{definition}{(Transitive Closure)}
  Let $e$ be an expression of type \xquery{node()*}.  The \emph{transitive
  closure} $e^{\xquery+}$ \emph{of} $e$ is 
  \begin{equation}
    e~\xquery{union}~e\xquery/e~\xquery{union}~e\xquery/e\xquery/e~\xquery{union}~\cdots
    \enskip,
    \label{xq:tc}
  \end{equation}
  if the resulting node sequence is finite.  Otherwise, $e^{\xquery+}$ 
  is \emph{undefined}.
\end{definition}
Given simple restrictions on $e$, see Section~\ref{sec:dist-in-xquery},
with the new IFP form $e^{\xquery+}$ is (`\xquery.' denotes the context
node): 
$$
\xquery{with \$x seeded by .\ recurse \$x/$e$}
\enskip.
$$

\begin{SaveVerbatim}[commandchars=\\\{\},codes={\catcode`$=3\catcode`\_=8}]{seed}
(SELECT prerequisite
 FROM   C
 WHERE  course = 'c1')
\end{SaveVerbatim}

\begin{SaveVerbatim}[baseline=c,commandchars=\\\{\},codes={\catcode`$=3\catcode`\_=8}]{body}
(SELECT C.prerequisite 
 FROM   P, C 
 WHERE  P.course\_code = C.course)
\end{SaveVerbatim}

\paragraph{IFP in SQL:1999.}
IFP has found its way into SQL in terms of the \sql{WITH RECURSIVE}
clause introduced by the ANSI/ISO SQL:1999 standard \cite{SQL99}.  To
exemplify, consider the table 
$\sql{C}(\sql{course},\sql{prerequisite})$ as a relational
representation of the curriculum XML data (Figure~\ref{fig:curri-data}).
The prerequisites $\sql{P}(\sql{course\_code})$ of the course with 
code~\sql{'c1'} then are:
$$
\renewcommand{\arraystretch}{1.1}
\begin{array}{ll}
  \sql{WITH RECURSIVE P(course\_code) AS} \\
  \quad
  \BUseVerbatim[baseline=c,baselinestretch=1.0]{seed}   
  & {\left.\vrule height 7mm width 0pt\right\}\text{\footnotesize seed}} \\
  ~~\sql{UNION ALL} \\
  \quad
  \BUseVerbatim[baseline=c,baselinestretch=1.0]{body} 
  & {\left.\vrule height 7mm width 0pt\right\}\text{\footnotesize body}} \\
  \sql{SELECT DISTINCT * FROM P;}
\end{array}
$$
Analogous to the XQuery variant, table~\sql{P} is seeded with the direct
prerequisites of course \sql{'c1'} before the join with table~\sql{C} in
the body is iterated to also add all indirect prerequisites until \sql{P}
does not grow further.

The SQL:1999 standard dictates quite rigid syntactical restrictions for
the \sql{WITH RECURSIVE} form (the body, in particular, must be
\emph{linear}: \sql{P} may occur only once in its \sql{FROM} clause).
We will return to this in Section~\ref{sec:syntactic-approx}
and~\ref{sec:related}.

\SubSection{Algorithms for IFP}

The semantics of the inflationary fixed point in XQuery, \emph{i.e.},
the specification of the node sequence $\var{res}_k$ of
Definition~\ref{def:ifp}, can be straightforwardly turned into an
iterative algorithm to compute IFP.  Figure~\ref{fig:naive_alg} shows
the resulting procedure, commonly referred to as \Naive in the 
literature \cite{rec_overview}.  In the $\textbf{do}\cdots\textbf{while}$
loop body, the procedure calls out to the recursion's payload function
$e_{\var{rec}}(\cdot)$ to determine the next portion of nodes that will
augment the current intermediate result.  Only if $e_{\var{rec}}(\cdot)$ cannot
contribute new nodes, the procedure returns the current $\var{res}$.

Since $\var{res}$ grows, this feeds the same nodes over and over again into
$e_{\var{rec}}(\cdot)$.  Dependent on the nature
of the payload, $e_{\var{rec}}(\cdot)$'s answer might include nodes
which we have seen before. Ultimately, \Naive risks to initiate a
substantial amount of redundant computation.

\begin{figure}
  \hskip-3mm\small
  \setlength{\subfigcapskip}{4pt}
  \subfigure[Algorithm \Naive]{%
  \label{fig:naive_alg}
  \begin{minipage}[t]{0.54\linewidth}
  \begin{algorithm}[H]
    $\var{res} \gets e_{\var{rec}}(e_{\var{seed}})$\;
    \phantom{\;}
    \Repeat{\var{res} \textnormal{grows}}{%
      \phantom{()\;}
      $\var{res} \gets e_{\var{rec}}(\var{res}) \mathbin{\xquery{union}} \var{res}$\;
    }
  \end{algorithm}
  \end{minipage}}
  \nolinebreak
  \vrule
  \nolinebreak
  \subfigure[Algorithm \DELTA]{%
  \label{fig:delta_alg}
  \begin{minipage}[t]{0.55\linewidth}
  \begin{algorithm}[H]
    $\var{res} \gets e_{\var{rec}}(e_{\var{seed}})$\;
    $\Delta \gets \var{res}$\;
    \Repeat{\var{res} \textnormal{grows}}{%
      $\Delta \gets e_{\var{rec}}(\Delta) \mathbin{\xquery{except}} \var{res}$\;
      $\var{res} \gets \Delta \mathbin{\xquery{union}} \var{res}$\;
    }
  \end{algorithm}
  \end{minipage}}
  \caption{Algorithms to evaluate the IFP of $e_{\var{rec}}$ given
    $e_{\var{seed}}$.  Result is $\var{res}$.}
  \label{fig:naive_delta_alg}
\end{figure}  

\smallskip\noindent A now folklore variation of \Naive is the \DELTA
algorithm~\cite{rec_delta} of Figure~\ref{fig:delta_alg}. In this
variant, the payload is invoked only for those nodes that have not
been encountered in earlier iterations: node sequence $\Delta$ is the
difference between $e_{\var{rec}}(\cdot)$'s last answer and the current
result $\var{res}$. In general, $e_{\var{rec}}(\cdot)$ will thus process
fewer nodes.

\begin{figure}
  \centering\small
    \begin{BVerbatim}[baselinestretch=0.9,
                    commandchars=\\\{\},codes={\catcode`$=3\catcode`\_=8}]
declare function delta\,(\$x,\$res) as node()* 
\{ let \$delta := rec\,(\$x) except \$res
  return if (empty\,(\$delta))  
         then \$res
         else delta\,(\$delta,\$delta union \$res)
\};
  \end{BVerbatim}
  \caption{An XQuery formulation of Delta.}
  \label{fig:delta-intro}
\end{figure}

\DELTA introduces a significant potential for performance improvement,
especially for large node sequences and computationally expensive
payloads (Section~\ref{sec:evaluation}). Figure~\ref{fig:delta-intro}
shows the corresponding XQuery user-defined function
\xquery{delta($\cdot$,$\cdot$)} which, for Example~\ref{ex:curriculum}
and thus Query~\ref{q:curri}, can
serve as a drop-in replacement for function
\xquery{fix($\cdot$)}---line~14 then needs to be replaced by
\xquery{return delta\,(rec\,(\$seed),())}.

\smallskip\noindent
Is this replacement of \xquery{fix($\cdot$)} by
\xquery{delta($\cdot$,$\cdot$)} \emph{always} a valid optimization?
For XQuery, the answer is \emph{no}.
\begin{example}
Consider the following expression:
\begin{equation}
\begin{BVerbatim}[baseline=c,
                  commandchars=\\\{\},codes={\catcode`$=3\catcode`\_=8}]
let \$seed := (<a/>,<b><c><d/></c></b>)
return with \$x seeded by \$seed
       recurse if (count\,(\$x/self::a))
               then \$x/* else () 
\end{BVerbatim}
\tag{Q2}
\label{q:count}
\end{equation}
Let $a$, $b$, $c$, and $d$ denote the tree fragments constructed by the
seed's subexpressions \xquery{<a/>}, \xquery{<b><c><d/></c></b>}, 
\xquery{<c><d/></c>}, and \xquery{<d/>}, respectively.  Thus,
\xquery{$b$/*\,is\,$c$} and \xquery{$c$/*\,is\,$d$}.

The table below illustrates the progress of the iterations performed by
algorithms \Naive and \DELTA.  While the former computes
\xquery{($a$,$b$,$c$,$d$)}, the latter returns \xquery{($a$,$b$,$c$)}.
\begin{center}
  \small
  \begin{tabular}{clll}
    \toprule 
    \textbf{Iteration} & \multicolumn{1}{c}{\bfseries\Naive} & 
                         \multicolumn{2}{c}{\bfseries\DELTA}
    \\
                       & \multicolumn{1}{c}{$\var{res}$} & 
                         \multicolumn{1}{c}{$\var{res}$} & 
                         \multicolumn{1}{c}{$\Delta$}
    \\
    \cmidrule(lr){1-1}\cmidrule(lr){2-2}\cmidrule(lr){3-4}
    0  & \xquery{($a$,$b$)}      & \xquery{($a$,$b$)}   & \xquery{($a$,$b$)}
    \\  
    1  & \xquery{($a$,$b$,$c$)}    & \xquery{($a$,$b$,$c$)} & \xquery{($c$)}
    \\ 
    2  & \xquery{($a$,$b$,$c$,$d$)}  & \xquery{($a$,$b$,$c$)} & \xquery{()}
    \\
    3  & \xquery{($a$,$b$,$c$,$d$)}  
    \\
    \bottomrule
  \end{tabular}
\end{center}
\end{example}
What then is an effective characterization of those payloads for which
\Naive may safely be traded for \DELTA?


\Section{Trading \Naive for \DELTA}
\label{sec:trading}

We will now see that a simple notion of \emph{distributivity} for XQuery
expressions suffices to let an XQuery processor safely switch to a more
efficient evaluation mode for \xquery{with \$x seeded by
$e_{\fun{seed}}$ recurse $e_{\fun{rec}}$}: whenever expression
$e_{\fun{rec}}$ is \emph{distributive} (in the sense defined below),
algorithm \DELTA (Figure~\ref{fig:delta_alg}) preserves the
desired IFP semantics. While the \emph{distributivity} property is
undecidable in general, we present two safe and effective approximations
of distributivity, one formulated on the level of XQuery language
syntax, and one cast in terms of an algebraic XQuery representation. The
algebraic approximation will turn out to be particularly simple and
uniform (Section~\ref{sec:dist_impl}).

\SubSection{Distributivity in XQuery}
\label{sec:dist-in-xquery}

Obviously, \DELTA computes the IFP for given expressions
$e_{\fun{seed}}$ and $e_{\fun{rec}}$ if the algorithm produces the same
result as \Naive on the same inputs. In particular, the algorithms
are equivalent if both yield equivalent intermediate $\fun{result}$
sequences in each iteration of their $\textbf{do}\cdots\textbf{while}$ 
loops.  

In its first loop iteration, \Naive yields
\xquery{$e_{\fun{rec}}(e_{\fun{rec}}(e_{\fun{seed}}))$ union
$e_{\fun{rec}}(e_{\fun{seed}})$} which is equivalent to \emph{Delta}'s
first intermediate result
\xquery{$(e_{\fun{rec}}(e_{\fun{rec}}(e_{\fun{seed}}))$ except
$e_{\fun{rec}}(e_{\fun{seed}}))$ union $e_{\fun{rec}}(e_{\fun{seed}})$}.
For the second and further iterations, an inductive proof can show the
equivalence of all subsequent intermediate $\fun{result}$ sequences, if
we may assume that, for two item sequences $X_1, X_2$, we have
\begin{equation} e_{\fun{rec}}(X_1 \mathbin{\xquery{union}} X_2)
\setequal e_{\fun{rec}}(X_1) \mathbin{\xquery{union}} e_{\fun{rec}}(X_2)
\enskip. \label{eq:distributive} \end{equation} For lack of space, we do
not reproduce the straightforward equational reasoning behind the proof
here but refer to \cite{afanasiev07}.

\bigskip Note how \eqref{eq:distributive} resembles the
\emph{distributivity property} of functions defined on sets. Such a
function $e$ is \emph{distributive} if, for all non-empty sets $X$, $e(X)
= \bigcup_{y \in X} e(\{ y \})$. This property suggests a
divide-and-conquer evaluation strategy in which $e$ is applied to
subsets (singletons) of $X$ only.  We define the corresponding
\emph{distributivity property for XQuery} as follows:
\begin{definition}{Distributivity property for XQuery.}
  Let $e$ be an XQuery expression in which variable \xquery{\$x} 
  may occur free.
  Expression $e$ is \emph{distributive for} \xquery{\$x} if, for
  any item sequence $X \not= \xquery{()}$ and fresh variable \xquery{\$y}, 
  \begin{equation}
  \xquery{for \$y in $X$ return $e(\xquery{\$y})$} 
  \quad\setequal\quad
  e(X) \enskip.
  \label{eq:def-distributive}
  \end{equation}
  \label{def:distributivity}
\end{definition}
In particular, Equality~\eqref{eq:distributive} is a straightforward
consequence if we know that the recursion body $e_{\var{rec}}$ is 
distributive for its free variable.  Overall, we arrive at
the following sufficient condition for the applicability of \DELTA:
\begin{theorem}
  Consider the expression
  \textnormal{\xquery{with \$x seeded by}}~$e_{\var{seed}}$ 
  \textnormal{\xquery{recurse}}~$e_{\var{rec}}$.
  If $e_{\var{rec}}$ is distributive for \textnormal{\xquery{\$x}}, 
  then algorithm
  \DELTA computes the IFP of $e_{\var{rec}}$ given $e_{\var{seed}}$.
  \label{th:delta-ifp}
\end{theorem}

\paragraph{XPath Location Steps.}
XPath location steps are a prevalent example of distributive expressions
in XQuery. Any expression of the form $e(\xquery{\$x}) =
\xquery{\$x/$s$}$ is distributive for \xquery{\$x} if the step
subexpression $s$ neither contains
\begin{inparaenum}[\itshape(i)]
\item free occurrences of \xquery{\$x}, nor
  \label{loc-step-1}
\item calls to \xquery{fn:position()} and \xquery{fn:last()} that refer
  to the context item sequence bound to \xquery{\$x}, nor
  \label{loc-step-2}
\item node constructors.
  \label{loc-step-3}
\end{inparaenum}
To see this, note that the XQuery~Core equivalent \cite{xquery_fs} of
\xquery{\$x/$s$} is \xquery{fs:ddo(for \$fs:dot in \$x return $s$)}, and
then rewrite the lhs of Equation~\eqref{eq:def-distributive} into its
rhs, using the definition of $\setequal$.

\paragraph{Regular XPath.} These observations about the distributivity
of XPath location steps extend to Regular~XPath
\cite{BtCRegPath:06} and
thus also make this XPath extension susceptible to \DELTA-based
evaluation. Since any Regular~XPath step subexpression $s$ is of the
form prescribed by {\itshape(\ref{loc-step-1})} to
{\itshape(\ref{loc-step-3})} above and Regular~XPath's transitive
closure $s^{\xquery{+}}$ is equivalently expressed as \xquery{with \$x
seeded by .\@ recurse \$x/$s$} (for the simple proof see
\cite{afanasiev07}), Theorem~\ref{th:delta-ifp} asserts that we may indeed
use algorithm \DELTA to evaluate $s^{\xquery{+}}$.

\bigskip\noindent
In contrast, expression $e(\xquery{\$x}) = \xquery{\$x[1]}$ is
\emph{not} distributive for \xquery{\$x} in general. With variable
\xquery{\$x} bound to the sequence \xquery{(<a/>,<b/>)}, \xquery{\$x[1]}
evaluates to \xquery{<a/>}, while \xquery{for \$y in \$x return \$y[1]}
yields \xquery{(<a/>,<b/>)}.  Effectively, this invalidates
Equation~\eqref{eq:def-distributive}.

\SubSection{Is Expression {\large$\emph{e}_{\emph{rec}}$} Distributive? \\
 (A Syntactic Approximation)}
\label{sec:syntactic-approx}

Whenever an XQuery processor plans the evaluation of \xquery{with \$x
seeded by $e_{\fun{seed}}$ recurse $e_{\fun{rec}}$}, knowing the answer
to ``\emph{Is $e_{\fun{rec}}$ distributive for
\textnormal{\xquery{\$x}}?}'' is particularly valuable: we may
legitimately expect \DELTA to be a significantly more efficient IFP
evaluation strategy than \Naive (Section~\ref{sec:evaluation} will
indeed make this evident). While, unfortunately, there is no complete
procedure to decide this question\footnote{If, for two arbitrary
expression $e_1$, $e_2$ in which \xquery{\$x} does not occur free, an
XQuery processor could assess whether \xquery{if (deep-equal($e_1$,$e_2$))
then \$x else \$x[1]} is distributive for \xquery{\$x}, it could also
decide the equivalence of $e_1$ and $e_2$ (which is impossible).}, still we
can safely approximate the answer. Here, we will present purely
syntactic, sufficient conditions for XQuery distributivity.
Section~\ref{sec:dist_impl} approaches the same challenge on an
algebraic level.

\smallskip\noindent Intuitively, we may \emph{not} apply a
divide-and-conquer evaluation strategy for an expression
$e(\xquery{\$x})$, if any subexpression of $e$ inspects the sequence
bound to \xquery{\$x} as a whole: $e$ is only evaluated after
\xquery{\$x} has been divided into individual items (see
Equation~\ref{eq:def-distributive}). Obvious examples of such
problematic subexpressions are \xquery{count(\$x)} and \xquery{\$x[1]},
but also the general comparison \xquery{\$x = 10} (that involves
existential quantification over the sequence bound to \xquery{\$x}).

Subexpressions whose value is \emph{independent} of \xquery{\$x}, on the
other hand, are distributive. The only exception of this rule are
XQuery's node constructors, \emph{e.g.}, \xquery{text\,\{$\cdot$\}},
which create new node identities upon each invocation. With \xquery{\$x}
bound to \xquery{(<a/>,<b/>)}, for example,
$$
\xquery{text\,\{\,"c"\,\}} 
\,\nsetequal\,
\xquery{for \$y in \$x return text\,\{\,"c"\,\}} \enskip,
$$
since the rhs will yield a sequence of two distinct text nodes.

\smallskip\noindent The inference rules of
Figure~\ref{fig:lixquery-distributivity} have been designed to implement
these considerations. The rules syntactically assess the
\emph{distributivity safety} $\ds{\xquery{\$x}}{e}$ of an arbitrary
LiXQuery \cite{LiXQuery} input expression $e$ by traversing $e$'s parse
tree in a bottom-up fashion. LiXQuery is a sublanguage of XQuery that
preserves Turing-completeness, removes all but the most basic types,
and excludes selected, rather esoteric, language features. LiXQuery's
simplification of the verbose XQuery syntax and semantics have been
designed to make LiXQuery ideal for investigations of interesting
language properties, yet allow findings to be transposed to full XQuery.

Rules~\ref{dist-for1} and~\ref{dist-for2} ensure that the recursion
variable \xquery{\$x} occurs either in the body $e_2$ or in the range
expression $e_1$ of a \xquery{for}-iteration but not both. This
coincides with the linearity constraint of SQL:1999. A similar remark applies
to Rules~\ref{dist-step1} and ~\ref{dist-step2} (in XQuery, the step
operator `\xquery/' essentially describes an iteration over a sequence
of type \xquery{node()*} \cite{xquery_fs}). Also note how
Rule~\ref{dist-funcall} recursively infers the distributivity of the
body of a called function if the recursion variable occurs
free in the function argument(s).

In our context, whenever the XQuery processor is able to infer
$\ds{\xquery{\$x}}{e}$ for an input expression $e$, then it is
guaranteed that $e$ is indeed distributive for \xquery{\$x}. The proof
of this implication, by induction on the syntactical structure of
$e$, is to be found in \cite{afanasiev07}.

\begin{figure*}
  \centering\small 
  $
  \begin{array}{@{}c@{}}
    \inferrule
      {\relax}
      {\ds{\xquery{\$x}}{c}}
    \distlabel{Const}{dist-const}
    \quad
    \inferrule
      {\relax}
      {\ds{\xquery{\$x}}{\xquery{\$}v}}
    \distlabel{Var}{dist-var}
    \quad
    \inferrule
      {\xquery{\$x} \notin \fv{e_1} \quad 
       \ds{\xquery{\$x}}{e_2}       \quad
       \ds{\xquery{\$x}}{e_3}}
      {\ds{\xquery{\$x}}{\xquery{if ($e_1$) then $e_2$ else $e_3$}}}
    \distlabel{If}{dist-if}
    \quad
    \inferrule
      {\oplus \in \{\xquery{,} , \xquery{|}\} \quad
       \ds{\xquery{\$x}}{e_1}                 \quad
       \ds{\xquery{\$x}}{e_2}}
      {\ds{\xquery{\$x}}{e_1 \oplus e_2}}
    \distlabel{Concat}{dist-concat}
    \\[5mm]
    \inferrule
      {\xquery{\$x} \notin \fv{e_1} \quad
       \ds{\xquery{\$x}}{e_2}}
      {\ds{\xquery{\$x}}{\xquery{for \$$v$ at \$$p$ in $e_1$ return $e_2$}}}
    \distlabel{For1}{dist-for1}
    \quad
    \inferrule
      {\ds{\xquery{\$x}}{e_1} \quad
       \xquery{\$x} \notin \fv{e_2}}
      {\ds{\xquery{\$x}}{\xquery{for \$$v$ in $e_1$ return $e_2$}}}
    \distlabel{For2}{dist-for2}
    \\[5mm]
    \inferrule
      {\xquery{\$x} \notin \fv{e_1} \quad
       \ds{\xquery{\$x}}{e_2}}
      {\ds{\xquery{\$x}}{\xquery{let \$$v$ := $e_1$ return $e_2$}}}
    \distlabel{Let1}{dist-let1}
    \quad
    \inferrule
      {\ds{\xquery{\$x}}{e_1}       \quad
       \xquery{\$x} \notin \fv{e_2} \quad
       \ds{\xquery{\$$v$}}{e_2}}
      {\ds{\xquery{\$x}}{\xquery{let \$$v$ := $e_1$ return $e_2$}}}
    \distlabel{Let2}{dist-let2}
    \\[5mm]
    \inferrule
      {\xquery{\$x} \notin \fv{e_1} \quad
       \ds{\xquery{\$x}}{c_i}_{i = 1\dots n+1}}
      {\ds{\xquery{\$x}}{%
        \begin{array}{@{}l@{~~}l@{}}
          \multicolumn{2}{l}{\xquery{typeswitch\,($e_1$)}} \\
          \quad \xquery{case $\tau_1$} & \xquery{return $c_1$} \\[-1ex]
          \multicolumn{2}{c}{$\vdots$} \\[-1ex]
          \quad \xquery{case $\tau_n$} & \xquery{return $c_n$} \\          
          \quad \xquery{default}       & \xquery{return $c_{n+1}$}
        \end{array}
      }}
    \distlabel{TypeSw}{dist-typesw}
    \qquad
    \begin{array}[t]{@{}l@{}}
    \inferrule
      {\xquery{\$x} \notin \fv{e_1} \quad
       \ds{\xquery{\$x}}{e_2}}
      {\ds{\xquery{\$x}}{\xquery{$e_1$/$e_2$}}}
    \distlabel{Step1}{dist-step1}
    \\[5mm]
    \inferrule
      {\ds{\xquery{\$x}}{e_1} \quad
       \xquery{\$x} \notin \fv{e_2}}
      {\ds{\xquery{\$x}}{\xquery{$e_1$/$e_2$}}}
    \distlabel{Step2}{dist-step2}
    \end{array}
    \\[20mm]
    \inferrule
      {\xquery{declare function $f$(\$$v_1$,$\dots$,\$$v_n$)\,\{\,$e_0$\,\}} \quad
       (\xquery{\$x} \in \fv{e_i} \Rightarrow 
       \ds{\xquery{\$x}}{e_i} \wedge \ds{\xquery{\$$v_i$}}{e_0})_{i=1\dots n}}
      {\ds{\xquery{\$x}}{\xquery{$f$($e_1$,$\dots$,$e_n$)}}}
    \distlabel{FunCall}{dist-funcall}
  \end{array}
  $
  \caption{Distributivity-safety $\ds{\xquery{\$x}}{\cdot}$:
    A syntactic approximation of the distributivity property
    for LiXQuery expressions.}
  \label{fig:lixquery-distributivity}
\end{figure*}

\paragraph{Distributivity Hints.} Still, the inference rules of
Figure~\ref{fig:lixquery-distributivity} can only check \emph{sufficient
syntactical conditions} for distributivity to hold. The processor
might thus actually miss distributive expressions and will fail to infer
$\ds{\xquery{\$x}}{\xquery{count(\$x) >= 1}}$, for example. However, it
is interesting to note that we can support the XQuery processor in its
distributivity assessment, since every distributive expression is
equivalent to a distributivity-safe expression:
\begin{list}{}{\leftmargin1em\rightmargin\leftmargin}
  \item
  \emph{If expression} $e(\xquery{\$x})$ \emph{is distributive for}
  \xquery{\$x}, then it is set-equal to \xquery{for \$y in \$x return
  $e(\xquery{\$y})$}, for which the rules of
  Figure~\ref{fig:lixquery-distributivity} will successfully infer
  distributivity safety $\ds{\xquery{\$x}}{\cdot}$.
\end{list}
This is a direct consequence of Rule~\ref{dist-for2}
(Figure~\ref{fig:lixquery-distributivity}) and
Definition~\ref{def:distributivity}. Thus, at the expense of a slight
query reformulation, we may provide a ``syntactic distributivity hint''
to the XQuery processor which effectively paves the way for IFP
evaluation via algorithm~\DELTA.


\Section{Distributivity and Relational XQuery}              
\label{sec:dist_impl}

In this section we will, literally, follow an alternative route to
decide the applicability of \DELTA for the evaluation of the IFP of an
XQuery expression $e_{\var{rec}}$. We leave syntax aside and instead
inspect \emph{relational algebraic code} that has been compiled for
$e_{\var{rec}}$: the equivalent algebraic representation of
$e_{\var{rec}}$ renders the check for the inherently algebraic
distributivity property particularly uniform and simple.

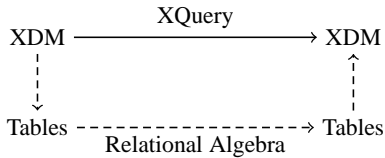
\begin{figure}
  \centering\small
  \begin{tikzpicture}[x=21mm,y=6mm,semithick]
    \draw 
      (-1,1) node (lu) {XDM}
      (1,1) node (ru) {XDM}
      (-1,-1) node (ld) {Tables}
      (1,-1) node (rd) {Tables}
      (0,1) node[above] {XQuery}
      (0,-1) node[below] {Relational Algebra}
    ;
    \draw[->,densely dashed] (lu) -- (ld);
    \draw[->,densely dashed] (rd) -- (ru);
    \draw[->] (lu) -- (ru);
    \draw[->,densely dashed] (ld) -- (rd);
  \end{tikzpicture}
\caption{Relational XQuery 
  (dashed path) 
  faithfully implements the XQuery semantics.}
\label{fig:XQ_relAlg}
\end{figure}

\paragraph{Relational XQuery.}
This alternative route is inspired by the \emph{Pathfinder}
project\footnote{\url{http://www.pathfinder-xquery.org/}} which fully
implements such a purely relational approach to XQuery.
\emph{Pathfinder} compiles instances of the XQuery
Data~Model (XDM) and XQuery expressions into relational tables
and algebraic plans over these tables, respectively, and thus follows
the dashed path in Figure~\ref{fig:XQ_relAlg}.  The translation strategy
built into the compiler has been carefully designed 
\begin{inparaenum}[(\itshape i)]
\item to faithfully preserve the XQuery semantics 
  (including compositionality, 
  node identity, iteration and sequence order), and
\item yield relational plans which exclusively rely on regular 
  relational query engine technology (no specific
  operators or index structures are required, in particular) 
  \cite{sql-mapping}.
\end{inparaenum}

The compiler emits a dialect of relational algebra that mimics the
capabilities of modern SQL query engines
(Table~\ref{tab:relational-operators}). Note that
the non-textbook operators, like $\varepsilon$ or $\scj$, merely are
macros representing ``micro plans'' composed of standard relational
operators: expanding $\scj_{\alpha\xquery{::}n}$ reveals $\Join_p$,
where $p$ is a conjunctive range predicate that realizes the semantics
of an XPath location step along axis $\alpha$ with node test $n$, for
example. The row numbering operator $\varrho$
directly compares with SQL:1999's \sql{ROW\_NUMBER}.  The plans operate
over relational encodings of XQuery item sequence held in flat (1NF)
tables with schema $\col{iter}|\col{pos}|\col{item}$. In these tables,
columns $\col{iter}$ and $\col{pos}$ are used to properly reflect
\xquery{for}-iteration and sequence order, respectively. Column
$\col{item}$ carries encodings of XQuery items, \emph{i.e.}, atomic
values or nodes.

Further details of Relational XQuery do not affect our present
discussion of distributivity or IFP evaluation and may be found in
\cite{sql-mapping}. In the following, let \miniaplan{$e$} denote the
algebraic plan that has been compiled for XQuery expression~$e$.

\begin{table}[t]
\centering\small
\begin{tabular}{@{}ll@{\,}c@{}}
\toprule
    Operator
  & Semantics 
  & Push? \\
    \cmidrule(r){1-1}\cmidrule(lr){2-2}\cmidrule{3-3}
    $\pi_{\col{a}_1:\col{b}_1, \dots, \col{a}_n:\col{b}_n}$
  & project onto col.s $\col{a}_i$, rename $\col{b}_i$ into $\col{a}_i$
  & $\odot$ 
  \\[-.1em]
    $\sigma_{\col{b}}$
  & select rows with column $\col{b} = \var{true}$
  & $\odot$ 
  \\[-.1em]
    $\Join_p$
  & join with predicate $p$
  & $\varoplus$ 
  \\[-.1em]
    $\times$
  & Cartesian product
  & $\varoplus$ 
  \\[-.1em]
    $\delta$
  & duplicate elimination (\sql{DISTINCT})
  & $-$
  \\[-.1em]
    $\cup$
  & union 
  & $\varoplus$ 
  \\[-.1em]
    $\setminus$
  & disjoint difference (\sql{EXCEPT ALL})
  & $-$
  \\[-.1em]
    $\texttt{count}_{\col{a}:/\col{b}}$
  & aggregates (group by $\col{b}$, result in $\col{a}$)
  & $-$
  \\[-.1em]
    $\circledcirc_{\col{a}:\left<\col{b}_1,\dots,\col{b}_n\right>}$
  & $n$-ary arithmetic/comparison operator $\circ$
  & $\odot$ 
  \\[-.1em]
    $\num_{\col{a}}$
  & unique row tagging (tag in $\col{a}$)
  & $\odot$ 
  \\[-.1em]
    $\varrho_{\col{a}:\left<\col{b}_1,\dots,\col{b}_n\right>/p}$
  & ordered row numbering (by $\col{b}_1,\dots,\col{b}_n$) 
  & $-$
  \\[-.1em]
    $\scj_{\alpha\xquery{::}n}$
  & XPath step join (axis $\alpha$, node test $n$)
  & $\odot$ 
  \\[-.1em]
    $\varepsilon,\tau,\dots$
  & node constructors
  & $-$
  \\[-.1em]
    $\mu, \mu^\Delta$
  & fixpoint operators
  & $\odot$ 
  \\[-.1em]
\bottomrule
\end{tabular}
\caption{Relational algebra dialect emitted by the 
  Pathfinder compiler.}
\label{tab:relational-operators}
\end{table}

\SubSection{Is Expression {\large$\emph{e}_{\emph{rec}}$} Distributive? \\
  (An Algebraic Account)}
\label{sec:algebraic-dist}

An occurrence of the new 
\xquery{with \$x seeded by $e_{\var{seed}}$ recurse $e_{\var{rec}}$}
form in a source XQuery expression will be 
\begin{wrapfigure}[8]{r}[3mm]{0mm}
  \centering\small
  \hskip-3mm
  \begin{tikzpicture}[x=3.5mm,y=7mm,inner sep=0.5pt]
    \draw
      (0,0) node (seed) {\waplan{$e_{\mskip-2mu\var{seed}}$}}
      (0,2) node (rec)  {\waplan{$e_{\var{rec}}$}}
      (0,1.1) node (in) {}
      (0,3) node (out) {}
      (0,3.5) node (exit) {}
    ;
    \draw[-o]
      (seed) -- (in)
    ;
    \draw[-o]
      (rec) -- (out)
    ;
    \draw[)-]
      (in) -- (rec)
    ;
    \draw
      (out) -- (exit)
    ;
    \draw[->,densely dashed,rounded corners=5pt]
      (out.north) -- ++(0,0.25) -- ++(-2,0) -- 
      node[midway,left] {\scriptsize\xquery{\$x}} 
      ++(0,-2.75) -- ++(1,0) node (arrow) {}
    ;
    \draw[densely dashed,rounded corners=5pt] 
      (arrow) -- ++(1,0) -- ++(0,0.25)
    ;
    \draw[snake=brace,raise snake=-1mm]
      (1.75,3) -- (1.75,0.75) node[midway,right] {\scriptsize$\mu$}
    ;
  \end{tikzpicture}
\end{wrapfigure}
compiled into a plan fragment as shown here on the right. Operator
$\mu$, the algebraic representation of algorithm \Naive
(Figure~\ref{fig:naive_alg}), iterates the evaluation of the algebraic
plan for $e_{\var{rec}}$ and feeds its output $\recout$ back to its
input $\recin$ until the IFP is reached. If we can guarantee that the
plan for $e_{\var{rec}}$ is distributive, we may safely trade $\mu$ for
its \DELTA-based variant $\mu^\Delta$ which, in general, will feed
significantly less items back in each iteration
(see Figure~\ref{fig:delta_alg} and Section~\ref{sec:evaluation}).

\begin{figure}
  \centering\small
  \subfigure[Is $e_{\var{rec}}$ distributive?]{%
  \label{fig:union-push-up}
  \begin{tikzpicture}[x=3.5mm,y=5mm,inner sep=0.5pt]
    \draw
      (0, 0.75) node (bef-rec)   {\waplan{$e_{\var{rec}}$}}
      (0,-0.5)  node (bef-u)     {$\cup$}
    ;
    \draw[)-] ( 1,-1.5) -- ( 1,-1);
    \draw[)-] (-1,-1.5) -- (-1,-1);
    \draw     ( 1,-1) -- (bef-u);
    \draw     (-1,-1) -- (bef-u);
    \draw     (bef-u) -- (bef-rec);
    \draw[-o] (bef-rec) -- (0,2);
    ;
    \path (3,0.25) node {$\overset{?}{=}$};
    ;
    \draw
      (5,-0.25) node (aft-rec1)   {\waplan{$e_{\var{rec}}$}}
      (8,-0.25) node (aft-rec2)   {\waplan{$e_{\var{rec}}$}}
      (6.5,1.2) node (aft-u)      {$\cup$}
    ;
    \draw[)-] (5,-1.5) -- (aft-rec1);
    \draw[)-] (8,-1.5) -- (aft-rec2);
    \draw     (aft-rec1) -- (5,0.45);
    \draw     (aft-rec2) -- (8,0.45);
    \draw     (5,0.45) -- (aft-u);
    \draw     (8,0.45) -- (aft-u);
    \draw[-o] (aft-u) -- (6.5,2);
  \end{tikzpicture}
  }
  \hfill
  \subfigure[Taking a \emph{big step}: Pushing $\cup$ through a plan template]{%
  \label{fig:big-step-push-up}
    \begin{tikzpicture}[x=3.5mm,y=5mm,inner sep=0.5pt]
    \draw
      (0,2)    node (bef-up)    {}
      (0,0.63) node (bef-op0)   {$\bigstep{}{5mm}{10mm}{}$}
      (0,-1)   node (bef-u0)    {$\cup$}
      (-1,-2)  node (bef-plan1) {\aplan{$e_1$}}
      (1,-2)   node (bef-plan2) {\aplan{$e_2$}}
    ;
    \draw
      (bef-up)  -- (bef-op0)
      (bef-op0) -- (bef-u0)
      (bef-u0)  -- (bef-plan1.north)
      (bef-u0)  -- (bef-plan2.north)
    ;
    \path (2.5,0) node {$\to$};
    ;
    \draw
      (6.5,2)   node (aft-up)    {}
      (6.5,1.5) node (aft-u0)    {$\cup$}
      (5,0)     node (aft-op0)   {$\bigstep{}{5mm}{10mm}{}$}
      (8,0)     node (aft-op1)   {$\bigstep{}{5mm}{10mm}{}$}
      (5,-2)    node (aft-plan1) {\aplan{$e_1$}}
      (8,-2)    node (aft-plan2) {\aplan{$e_2$}}
    ;
    \draw
      (aft-up)  -- (aft-u0)
      (aft-u0)  -- (aft-op0.north)
      (aft-u0)  -- (aft-op1.north)
      (aft-op0) -- (aft-plan1.north)
      (aft-op1) -- (aft-plan2.north)
    ;
  \end{tikzpicture}
  }
  \caption{Algebraic distributivity assessment.}
  \label{fig:push-up}
\end{figure}
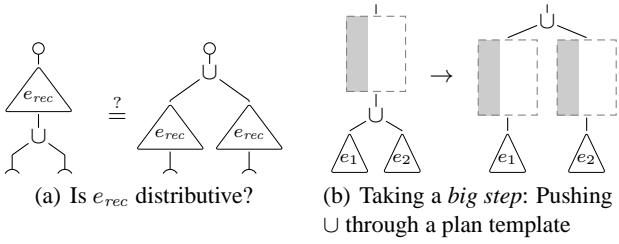

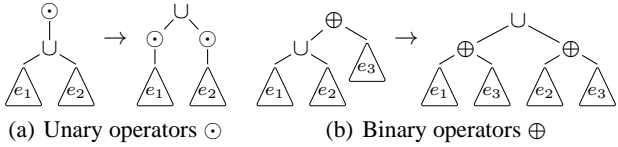
\begin{figure}
  \centering\small
  \subfigure[Unary operators $\odot$]{%
  \label{fig:u_trans_a}
  \begin{tikzpicture}[x=3.5mm,y=5mm,inner sep=0.5pt]
    \draw
      (0,0)   node (bef-op0)   {$\odot$}
      (0,-1)  node (bef-u0)    {$\cup$}
      (-1,-2) node (bef-plan1) {\aplan{$e_1$}}
      (1,-2)  node (bef-plan2) {\aplan{$e_2$}}
    ;
    \draw
      (bef-op0) -- (bef-u0)
      (bef-u0)  -- (bef-plan1.north)
      (bef-u0)  -- (bef-plan2.north)
    ;
    \path (2.5,-0.75) node {$\to$};
    ;
    \draw
      (5,0)     node (aft-u0)    {$\cup$}
      (4,-0.75) node (aft-op0)   {$\odot$}
      (6,-0.75) node (aft-op1)   {$\odot$}
      (4,-2)    node (aft-plan1) {\aplan{$e_1$}}
      (6,-2)    node (aft-plan2) {\aplan{$e_2$}}
    ;
    \draw
      (aft-u0)  -- (aft-op0)
      (aft-u0)  -- (aft-op1)
      (aft-op0) -- (aft-plan1.north)
      (aft-op1) -- (aft-plan2.north)
    ;
  \end{tikzpicture}
  }
  \hfill
  \subfigure[Binary operators $\varoplus$]{%
  \label{fig:u_trans_b}
  \begin{tikzpicture}[x=3.5mm,y=5mm,inner sep=0.5pt]
    \draw
      (0,-0.25)         node (bef-op0)   {$\varoplus$}
      (-1.25,-1) node (bef-u0)    {$\cup$}
      (-2.25,-2) node (bef-plan1) {\aplan{$e_1$}}
      (-0.25,-2) node (bef-plan2) {\aplan{$e_2$}}
      (1.25,-1.35)   node (bef-plan3) {\aplan{$e_3$}}
    ;
    \draw
      (bef-op0) -- (bef-u0)
      (bef-op0) -- (bef-plan3.north)
      (bef-u0)  -- (bef-plan1.north)
      (bef-u0)  -- (bef-plan2.north)
    ;
    \path (2.75,-0.75) node {$\to$};
    ;
    \draw
      (7,-0.25)     node (aft-u0)    {$\cup$}
      (5,-1) node (aft-op0)   {$\varoplus$}
      (9,-1) node (aft-op1)   {$\varoplus$}      
      (4,-2) node (aft-plan1) {\aplan{$e_1$}}
      (6,-2) node (aft-plan2) {\aplan{$e_3$}}
      (8,-2) node (aft-plan3) {\aplan{$e_2$}}
      (10,-2) node (aft-plan4) {\aplan{$e_3$}}      
    ;
    \draw
      (aft-u0)  -- (aft-op0)
      (aft-u0)  -- (aft-op1)
      (aft-op0) -- (aft-plan1.north)
      (aft-op0) -- (aft-plan2.north)
      (aft-op1) -- (aft-plan3.north)
      (aft-op1) -- (aft-plan4.north)
    ;
  \end{tikzpicture}
  }
\caption{Pushing $\cup$ through unary ($\odot$) and binary ($\varoplus$) 
  operators.}
\label{fig:u_trans}
\end{figure}

\smallskip\noindent
In this algebraic setting, if the recursion body $e_{\var{rec}}$ is
distributive, its relational plan will satisfy the equality shown in
Figure~\ref{fig:union-push-up}. This equality is the algebraic
expression of a divide-and-conquer evaluation strategy for
$e_{\var{rec}}$ (Section~\ref{sec:dist-in-xquery}): evaluating
$e_{\var{rec}}$ over a composite input 
(lhs, 
\begin{tikzpicture}[baseline=0.6ex,x=1.25mm,y=2mm,inner sep=1pt]
    \draw (3,1.5) node (union) {$\cup$};
    \draw[)-] (1,0) -- (1,1);
    \draw[)-] (5,0) -- (5,1);
    \draw
       (1,1) -- (union.west)
       (5,1) -- (union.east)
    ;
    \useasboundingbox (0,0) rectangle (6,2);        
\end{tikzpicture}) 
yields the same result as the union of the evaluation of $e_{\var{rec}}$
over a partitioned input (rhs). Effectively, the union operator $\cup$
has been completely pushed up through all branches of the DAG-shaped
algebraic plan for $e_{\var{rec}}$. Zooming in from the plan to the
operator level, Figure~\ref{fig:u_trans} depicts how $\cup$ is pushed up
through unary ($\odot$) and binary ($\varoplus$) operators. Column
`Push{?}' of Table~\ref{tab:relational-operators} indicates whether
$\cup$ may indeed be validly pushed through a given operator. Note that
this push through is prohibited by exactly those operators that require
to consume their \emph{complete} input to produce the result. This
affects, \emph{e.g.}, aggregates, difference, and the row numbering
operator. As before, the occurrence of node constructors renders
$e_{\var{rec}}$ non-distributive.

\smallskip\noindent Because our primary goal is distributivity assessment (as
opposed to query \emph{evaluation}---but see
Section~\ref{sec:evaluation}), we may actually employ simplified
variants of $e_{\var{rec}}$ in this context. In particular, since the
definition of distributivity disregards duplicates and order
(Definition~\ref{def:distributivity}), the compiler may choose to remove
code from $e_{\var{rec}}$ that is used to eliminate duplicate nodes
after XPath location steps as well as omit those parts of the plan that
realize the proper XQuery order semantics \cite{exrquy}.

\begin{figure}
  \centering\small
  \subfigure[$e_{\var{rec}}$ of Query~\ref{q:curri}]{%
  \label{fig:curri_alg}
      \centering\small
      \begin{tikzpicture}[x=19mm,y=6mm,inner sep=1pt]
        \draw 
          (0,-1) node (in) {}
          (0,0)  node (pi0)
                 { $\pi_{\col{iter},\col{item}}$ }
          (0,1)  node (num0)
                 { $\num_{\col{inner}}$ }
          (0,4.5)  node (pi4)
                 { $\pi_{%
                    \renewcommand{\arraystretch}{0.5}             
                    \begin{array}{@{}l@{}}
                    \scriptstyle
                    \col{inner}, \\
                    \scriptstyle                 
                    \col{outer}:\col{iter}
                    \end{array}}
                  $ 
                 }
          (1,2)  node (pi1)
                 { $\pi_{\col{iter:inner},\col{item}}$ }
          (1,3.25)  node (scj0)       
                 { $\underset{%
                      \renewcommand{\arraystretch}{0.5}
                      \begin{array}{@{}c@{}}
                         \scriptstyle
                         \xquery{child::} \\
                         \scriptstyle
                         \xquery{prerequisites}
                      \end{array}}{\scj}$ }
          (1,4.85)  node (scj1)       
                 { $\underset{%
                      \renewcommand{\arraystretch}{0.5}
                      \begin{array}{@{}c@{}}
                         \scriptstyle
                         \xquery{child::} \\
                         \scriptstyle
                         \xquery{pre\_code}
                      \end{array}}{\scj}$ }
          (2,4.85)  node (idref) 
                 { $
                   \begin{littbl}
                     \begin{array}{|c|c|}
                       \colhd{\scriptstyle id} & 
                       \colhd{\scriptstyle ref} \\[-1ex]
                       \vdots       & \vdots       \\
                       \hline
                     \end{array}
                   \end{littbl}
                   $ }
          (1.5,6)  node (join0)
                 { $\underset{\col{item} = \col{id}}{\Join}$ }
          (1.5,7)  node (pi2)
                 { $\pi_{\col{iter},\col{item}:\col{ref}}$ }
          (0,8)  node (join1)
                 { $\underset{\col{inner} = \col{iter}}{\Join}$ }
          (0,9)  node (pi3)
                 { $\pi_{\col{iter}:\col{outer},\col{item}}$ }
          (0,10) node (out)  {}
        ;
        \draw[)-]
          (in) -- (pi0)
        ;
        \draw
          (pi0) -- (num0)
          (num0) -- (pi4)
          (num0) -- (1,1.5) -- (pi1)
          (pi4) -- (join1)
          (join1) -- (pi3)
          (pi1) -- (scj0)
          (scj0) -- (scj1)
          (scj1) -- (1,5.75) -- (join0)
          (idref) -- (2,5.75) -- (join0)
          (join0) -- (pi2)
          (pi2) -- (1.5,7.5) -- (join1)
        ;
        \draw[-o]
          (pi3) -- (out)
        ;
        \draw
          (0.875,4.6) node
                { $\bigstep{loop}{48mm}{47mm}{}$ }
          (0.93,4.05) node
                { $\bigstep{step}{15mm}{18mm}{}$ }
        ;
      \end{tikzpicture}
  }
  \hfill
  \subfigure[$e_{\var{rec}}$ of Query~\ref{q:count}]{%
  \label{fig:count_alg}
      \centering\small
      \begin{tikzpicture}[x=8mm,y=6mm,inner sep=1pt]
        \draw
          (0,-2)  node (in) {}
          (0,-1)  node (pi-1)
                  { $\pi_{\col{iter},\col{item}}$ }
          (0,0)   node[inner sep=0pt,outer sep=0pt] (split) {}
          (1,1.5) node (scj0)
                  { $\underset{\xquery{self::a}}{\scj}$ }
          (1,2.5) node (pi0)      
                  { $\pi_{\col{iter}}$ } 
          (1,3.5) node (count0)
                  { $\xquery{count}_{\col{item}/\col{iter}}$ }
          (0,5)   node (join0)
                  { $\underset{\col{iter}}{\Join}$ }
          (0,6.25) node (scj1)
                  { $\underset{\xquery{child::*}}{\scj}$ }
          (0,7.5) node (out) {}
          (1,4.5) node[outer sep=3pt] (ghost) {}
        ;
        \draw
          (0.875,1.5) node
                  { $\bigstep{step}{1cm}{7mm}{}$ }
          (-0.125,6.25) node
                  { $\bigstep{step}{1cm}{7mm}{}$ }
        ;
        \draw[)-] 
          (in) -- (pi-1)
        ;
        \draw 
          (pi-1) -- (0,0)
          (0,0) -- (1,0.5) -- (scj0)
          (scj0) -- (pi0)
          (pi0) -- (count0)
          (0,0) -- (-1,0.5) -- (-1,4.5) -- (join0)
          (join0) -- (scj1)
          (ghost) -- (join0)
        ;  
        \draw[densely dotted] 
          (count0) -- (ghost) 
        ;
        \draw[-o]
          (scj1) -- (out)
        ;
      \end{tikzpicture}
  }
\caption{Relational representations of the recursion bodies 
  $e_{\var{rec}}$ of Queries~\ref{q:curri} and~\ref{q:count}.}
\label{fig:plans}
\end{figure}
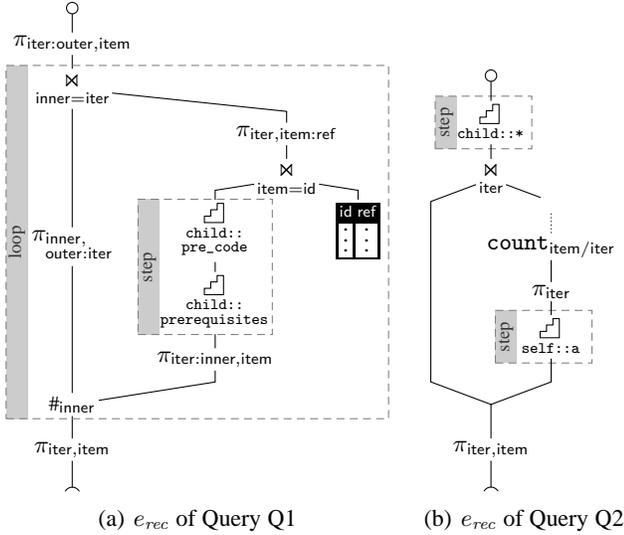

Further, the plans generated by the XQuery compiler typically contain
numerous instantiations of \emph{plan templates}, closed plan fragments
with single entry and exit points (enclosed by
\smash{\tikz[baseline=-0.3ex]\draw (0,0) node
{$\bigstep{}{3ex}{2ex}{}$};} in Figure~\ref{fig:plans}). These templates
embody algebraic implementations of basic XQuery constructs,
\emph{e.g.}, the semantics of \xquery{for}-iteration or XPath location
steps. Assessing the distributivity of such plan templates is a one time
effort. Once this has been done, whenever a distributive template is
encountered, the $\cup$ push up process may disregard the template's
contents and instead perform a single \emph{big step} across the template
(see Figure~\ref{fig:big-step-push-up}).

\smallskip\noindent
For the XQuery processor, this suggests the following simple procedure
as a replacement for $\ds{\xquery{\$x}}{\cdot}$
(Section~\ref{sec:syntactic-approx})
to assess the distributivity of $e_{\var{rec}}$:

\medskip\noindent
{
\setlength{\algomargin}{1ex}
\begin{algorithm}[H]
  Start with the algebraic plan for $e_{\var{rec}}$ with 
  its input $\recin$ replaced by 
  \begin{tikzpicture}[baseline=0.6ex,x=1.25mm,y=2mm,inner sep=1pt]
      \draw (3,1.5) node (union) {$\cup$};
      \draw[)-] (1,0) -- (1,1);
      \draw[)-] (5,0) -- (5,1);
      \draw
         (1,1) -- (union.west)
         (5,1) -- (union.east)
      ;
      \useasboundingbox (0,0) rectangle (6,2);        
  \end{tikzpicture}\;
  \While{\textnormal{not all $\cup$ have reached $\recout$}}{%
    Perform a \emph{big step} or push $\cup$ up through its parent operator,
    if possible. 
    Otherwise \Return{\var{false}}\;
  }
  \Return{\var{true}}\;
\end{algorithm}}

\medskip\noindent
Figure~\ref{fig:plans} depicts the algebraic representations of the
recursion bodies of the Queries~\ref{q:curri} and~\ref{q:count}
(Section~\ref{sec:IFP}). For Query~\ref{q:curri}, to push $\cup$ through
from $\recin$ to $\recout$, the distributivity check will succeed after
it has performed two steps across the two peripheral projections plus
one intermediate \emph{big step} across the \xquery{for}-iteration that
implements the semantics of the $\xquery{\$x/id($\cdot$)}$ lookup. For
Query~\ref{q:count}, $\cup$ will be pushed through
$\pi_{\col{iter},\col{item}}$ and then upwards the two branches of the
DAG-shaped plan. In the right branch, the aggregate
$\xquery{count}_{\col{item}/\col{iter}}$ blocks the process
(Table~\ref{tab:relational-operators}) which indicates that the processor
may \emph{not} use algorithm~\DELTA (or the $\mu^{\Delta}$ variant of
the fixed point operator) to evaluate Query~\ref{q:count}.

\paragraph{Algebraic vs.\ Syntactic Approximation.}
Compared to the syntactic approximation $\ds{\!}{\cdot}$,
this algebraic account of distributivity draws its conciseness from the
fact that the rather involved XQuery semantics and substantial number of
built-in functions nevertheless map to a small number of algebraic
primitives (given suitable relational encodings of the XDM). Further,
for these primitives, the algebraic distributivity property is readily
decided.

To make this point, consider this slight yet equivalent variation of 
Query~\ref{q:curri} in which variable \xquery{\$x} now occurs free in the
argument of function \xquery{id($\cdot$)}:
$$
\begin{BVerbatim}[baseline=c,commandchars=\\\{\},codes={\catcode`$=3\catcode`\_=
8}]
with \$x seeded by doc\,("curriculum.xml")
                    /course[@code="c1"]
recurse id\,(\$x/prerequisites/pre\_code) \enskip\textnormal{.}
\end{BVerbatim}
$$
If we unfold the implementation of the XQuery built-in function
\xquery{id($\cdot$)} (effectively, this expansion is performed when
Rule~\ref{dist-funcall} recursively invokes $\ds{\xquery{\$x}}{\cdot}$ to assess the
distributivity of the function body of \xquery{id($\cdot$)}), we obtain
\begin{equation*}
\begin{BVerbatim}[commandchars=\\\{\},codes={\catcode`$=3\catcode`\_=8}]
with \$x seeded by doc("curriculum.xml")
                    /course[@code="c1"]
recurse 
  for \$c in doc("curriculum.xml")/course
  where \$c/@code = \$x/prerequisite/pre\_code
  return \$c \enskip\textnormal{.}
\end{BVerbatim}
\end{equation*}
The syntactic approximation will flag the recursion body as
non-distributive because of the general comparison (\xquery{=}) in the
\xquery{where} clause (Section~\ref{sec:syntactic-approx}). 
While the algebraic approach would be unaffected by the variation,
the rule set
of Figure~\ref{fig:lixquery-distributivity} would need a
specific rule for \xquery{id($\cdot$)} to be able to infer its
actual distributivity.

\Section{Practical Impact of Distributivity and \DELTA}
\label{sec:evaluation}


\begin{table*}
  \centering\small
  \setlength{\tabcolsep}{1.6mm}
\begin{tabular}{lcrrcrrcrrcr}
\toprule
 \multicolumn{1}{l}{\textbf{Query}} && 
 \multicolumn{2}{c}{\textbf{\emph{MonetDB/XQuery}}}   &&
 \multicolumn{2}{c}{\textbf{\emph{Saxon-SA~8.9}}}     && 
 \multicolumn{2}{c}{\textbf{Total \# of Nodes Fed Back}} && 
 \multicolumn{1}{c}{\textbf{Recursion}} \\
 &&
 \multicolumn{1}{c}{\Naive} & \multicolumn{1}{c}{\DELTA} &&
 \multicolumn{1}{c}{\Naive} & \multicolumn{1}{c}{\DELTA} &&
 \multicolumn{1}{c}{\Naive} & \multicolumn{1}{c}{\DELTA} && 
 \multicolumn{1}{c}{\textbf{Depth}} \\
 \cmidrule(lr){1-1}\cmidrule(lr){3-4}\cmidrule(lr){6-7}\cmidrule(lr){9-10}\cmidrule(lr){12-12}
 Bidder network (small)     &&      362 ms &      165 ms &&    2,307 ms &     1,872 ms &&      40,254 &     9,319 && 10  \\
 Bidder network (medium)    &&    5,010 ms &    1,995 ms &&   15,027 ms &     7,284 ms &&     683,225 &   122,532 && 16  \\
 Bidder network (large)     &&   40,785 ms &   13,805 ms &&  123,316 ms &    52,436 ms &&   5,694,390 &   961,356 && 15  \\
 Bidder network (huge)      &&    9 m 46 s &  176,890 ms &&   32 m 40 s &    12 m 04 s &&  87,528,919 & 9,799,342 && 24  \\
[1ex]                                                                                                    
 Romeo and Juliet    &&    6,795 ms &    1,260 ms &&    1,150 ms &       818 ms &&      37,841 &     5,638 && 33  \\
[1ex]                                                                                                    
 Curriculum (medium) &&      183 ms &      135 ms &&    1,308 ms &     1,040 ms &&      12,301 &     3,044 && 18  \\
 Curriculum (large)  &&    1,466 ms &      646 ms &&    3,485 ms &     2,176 ms &&     127,992 &    19,780 && 35  \\
[1ex]                                                                                                    
 Hospital (medium)   &&      734 ms &      497 ms &&    1,301 ms &     1,290 ms &&      99,381 &    50,000 &&  5  \\
\bottomrule
\end{tabular}
\caption{Na\"ive vs.\ Delta: Comparison of query evaluation times and total number of nodes fed back.}
\label{tab:experiments}
\end{table*} 

Recasting a recursive XQuery query as an inflationary fixed point
computation imposes restrictions. Such recasting, however, also puts the
query processor into control since the applicability of a promising
optimization, trading \Naive for \DELTA, becomes effectively decidable.
This section provides the evidence that significant gains can indeed be
realized, much like in the relational domain.

To quantify the impact, we implemented the two fixed point operator
variants $\mu$ and $\mu^{\Delta}$ (Section~\ref{sec:algebraic-dist}) in
\emph{MonetDB/XQuery~0.18} \cite{monetdb-xquery}, an efficient and
scalable XQuery processor that consequently implements the
Relational~XQuery approach (Section~\ref{sec:dist_impl}).  
Its algebraic compiler front-end \emph{Pathfinder}  has
been enhanced 
\begin{inparaenum}[(\itshape i)]
\item to process the syntactic form
  $\xquery{with}\cdots\xquery{seeded by}\cdots\xquery{recurse}$, and
\item to implement the algebraic distributivity check.  
\end{inparaenum}
All queries in this section were recognized as being distributive by
\emph{Pathfinder}.  
To demonstrate that any XQuery processor can benefit from optimized IFP
evaluation in the presence of distributivity, we also performed the
transition from \Naive to \DELTA on the XQuery source level and let
\emph{Saxon-SA~8.9} \cite{saxon} process the resulting user-defined
recursive queries (cf.\ Figures~\ref{fig:fix-intro}
and~\ref{fig:delta-intro}). All experiments were conducted on a
Linux-based host (64\,bit), with two 3.2\,GHz
Intel~Xeon$^{\text{\textregistered}}$ CPUs, 8\,GB of primary and 280\,GB
SCSI disk-based secondary memory.

Table~\ref{tab:experiments} summarizes our observations for four query
types, chosen to inspect the systems' behavior for growing input XML
instance sizes and varying result sizes at each recursion level (the
maximum recursion depth ranged from~5 to~33).

\begin{figure}
\centering\small
\begin{BVerbatim}[baselinestretch=0.9,
                  commandchars=\\\{\},codes={\catcode`$=3\catcode`\_=8}]
declare variable \$doc := doc("auction.xml");

declare function bidder\,(\$in as node()*) as node()* 
\{ for \$id in \$in/@id
  let \$b := \$doc//open\_auction[seller/@person = \$id]
              /bidder/personref
  return \$doc//people/person[@id = \$b/@person]
\};

for \$p in \$doc//people/person
return <person>
        \{ \$p/@id \}
        \{ data ((with \$x seeded by \$p
                 recurse bidder\,(\$x))/@id) \}
       </person>
\end{BVerbatim}
\caption{XMark bidder network query.}
\label{q:network}
\end{figure}

\paragraph{XMark Bidder Network.} To assess scalability, we computed a
bidder network---recursively connecting the sellers and bidders of
auctions (Figure~\ref{q:network})---over XMark \cite{xmark} XML data of
increasing size (from scale factor~0.01, small, to~0.33, huge). If
\DELTA is used to compute the IFP of this network, \emph{MonetDB/XQuery}
(2.2~to~3.3 times faster) as well as \emph{Saxon} (1.2~to~2.7 times
faster) benefit significantly. Most importantly, note that the number of
nodes in the network grows quadratically with the input document size.
Algorithm \DELTA feeds significantly less nodes back in each recursion
level which positively impacts the complexity of the value-based join
inside recursion payload \xquery{bidder($\cdot$)}: for the huge network,
\DELTA exactly feeds those 10~million nodes into
\xquery{bidder($\cdot$)} that make up the result---\Naive repeatedly
revisits intermediate results and processes 9~times as many nodes.
  
\paragraph{Romeo and Juliet Dialogs.} Far less nodes are processed by
a recursive expression that queries XML markup of Shakespeare's Romeo
and
Juliet\footnote{\url{http://www.ibiblio.org/xml/examples/shakespeare/}}
to determine the maximum length of any uninterrupted dialog. Seeded with
\xquery{SPEECH} element nodes, each level of the recursion expands the
currently considered dialog sequences by a single \xquery{SPEECH} node
given that the associated \xquery{SPEAKER}s are found to alternate
(horizontal structural recursion along the \xquery{following-sibling}
axis---we do not reproduce the query here for space reasons.) Although
the recursion is shallow (depth~6 on average),
Table~\ref{tab:experiments} shows how both, \emph{MonetDB/XQuery} and
\emph{Saxon}, completed evaluation up to 5~times faster because the
query had been specified in a distributive fashion.

\paragraph{Transitive Closures.} Two more queries, taken directly from
related work \cite{xlinkit,smoqe}, compute transitive closure problems 
(we generated the data instances with the help of ToXgene \cite{toxgene}).
The first query implements a consistency check over the curriculum data
(cf.\ Figure~\ref{fig:curri-data}) and finds courses that are among their
own prerequisites (Rule~5 in the Curriculum Case Study in Appendix~B of
\cite{xlinkit}). Much like for the bidder network query, the larger the
query input (medium instance: 800~courses, large: 4,000~courses), the
better \emph{MonetDB/XQuery} and \emph{Saxon} exploited \DELTA.

The last query in the experiment explores 50,000 hospital patient
records to investigate a hereditary disease \cite{smoqe}. In this case,
the recursion follows the hierarchical structure of the XML input (from
patient to parents), recursing into subtrees of a maximum depth of~5.
Again, \DELTA makes a notable difference even for this computationally
rather ``light'' query.  

We believe that this renders this particular controlled form of XQuery
recursion and its associated distributivity notion attractive, even for
processors that do not implement a dedicated fixed point operator (like
\emph{Saxon}).

\Section{More Related Work}
\label{sec:related}
%
%

Bringing adequate support for recursion to XQuery is a core research
matter on various levels of the language. While the efficient evaluation
of the recursive XPath axes (\emph{e.g.}, \xquery{descendant} or
\xquery{ancestor}) is well understood by now
\cite{structural_join,scj}, the optimization of recursive
user-defined functions has been found to be tractable only in the
presence of restrictions: \cite{park02,grinev04} propose exhaustive
inlining of functions but require that functions are \emph{structurally}
recursive (use axes \xquery{child} and \xquery{descendant} to navigate
into subtrees only) over \emph{acyclic} schemata to guarantee that
inlining terminates. Note that, beyond inlining, this type of recursion
does not come packaged with an effective optimization hook comparable to
what the inflationary fixed point offers.

The distinguished use case for inflationary fixed point computation is
transitive closure. This is also reflected by the advent of XPath
dialects like Regular~XPath \cite{BtCRegPath:06} and
the inclusion of a dedicated \xquery{dyn:closure($\cdot$)} construct in
the EXSLT function library \cite{exslt}. We have seen applications in
Section~\ref{sec:evaluation} \cite{xlinkit,smoqe} and recent work on data
integration and XML~views adds to this \cite{fan07}.

\smallskip\noindent
In the domain of relational query languages, \Naive is the most
widely described algorithmic account of the inflationary fixed point
operator \cite{rec_overview}.  Its optimized \DELTA variant,
in focus since the 1980's, has been coined \emph{delta iteration}
\cite{rec_delta}, \emph{semi-\naive} \cite{rec_overview}, or
\emph{wavefront} \cite{rec_wavefront} strategy in earlier work.

Since our work rests on the adaption of these original ideas to the
XQuery Data Model and language, the large ``relational body'' of work in
this area should be directly transferable, even more so in the
Relational~XQuery context.  In particular, optimization techniques like
\emph{Magic Set} rewriting \cite{magic_set} should apply (this has not
been investigated in the present paper).

\smallskip\noindent The adoption of inflationary fixed point semantics by
Datalog and SQL:1999 with its \sql{WITH RECURSIVE} clause
(Section~\ref{sec:IFP}) led to investigations of the applicability of
\DELTA for these recursive relational query languages. For stratified
Datalog programs \cite{dbbook}, \DELTA is applicable in \emph{all}
cases: positive Datalog maps onto the distributive operators of 
relational algebra ($\pi$, $\sigma$, $\Join$, $\cup$, $\cap$) while
stratification yields partial applications of the difference
operator $x \setminus R$ in which $R$ is fixed ($f(x) = x \setminus R$ is 
distributive).

SQL:1999, on the other hand, imposes rigid \emph{syntactical}
restrictions \cite{SQL99} on the iterative fullselect (recursion body)
inside \sql{WITH RECURSIVE} that make \DELTA applicable: grouping,
ordering, usage of column functions (aggregates), and nested subqueries
are ruled out, as are repeated references to the virtual table computed
by the recursion. Replacing this coarse syntactic check by an algebraic
distributivity assessment (Section~\ref{sec:dist_impl}) would render a
larger class of queries admissible for efficient fixed point
computation.

%
%
%
%
%
%

\Section{Wrap-Up}
\label{sec:wrap-up}

This paper may be read in two ways:
\begin{asparaenum}[\itshape(i)]
\item As a proposal to add an inflationary fixed point construct, along
  the lines of $\xquery{with}\cdots\xquery{seeded
  by}\cdots\xquery{recurse}$, to XQuery (this has actually been discussed
  by the W3C~XQuery working group in the very early XQuery days of
  2001\footnote{\url{http://www.w3.org/TR/2001/WD-query-semantics-20010607/}
  (Issue~0008).} 
  but then dismissed because the group
  aimed for a first-order language design at that time).
  \label{read-as-1}
\item As a guideline for query authors as well XQuery processor designers
  to check for and then exploit distributivity during the evaluation of
  recursive queries.
  \label{read-as-2}
\end{asparaenum}

We have seen how such distributivity checks can be used to safely unlock
the optimization potential, namely algorithm~\DELTA, that comes tightly
coupled with the inflationary fixed point semantics.
\emph{MonetDB/XQuery} implements this distributivity check on the
algebraic level and significantly benefits whenever the \DELTA-based
operator $\mu^{\Delta}$ may be used for fixpoint computation. Even if
the approach is realized on the coarser syntactic level \emph{on top of}
an existing XQuery processor, feeding back less nodes in each recursion
level yields substantial performance improvements.

\smallskip\noindent Remember that the distributivity notion suggests a
divide-and-conquer evaluation strategy
(Section~\ref{sec:dist-in-xquery}) in which parts of a computation may
be performed independently (before a merge step forms the final result).
Beyond recursion, this may lead to improved XQuery compilation
strategies for back-ends that can exploit such independence, \emph{e.g},
set-oriented relational query processors (cf.\ loop-lifting
\cite{sql-mapping}) as well as parallel or distributed execution
platforms.

\paragraph{Acknowledgments.} Loredana Afanasiev is supported by
the Netherlands Organization for Scientific Research (NWO) under 
project number~612.000.207.  Jan Rittinger is supported by the German Research
Foundation (DFG) under grant GR~2036/2-1. We thank Massimo Franceschet
for input in an early stage of this work.

\bibliographystyle{latex8}
\renewcommand{\baselinestretch}{0.9}
\bibliography{references}

\end{document}